# Treating random sequential addition via the replica method

Ryan B. Jadrich,[1,2,a)] Beth A. Lindquist,[1,2] and Thomas M. Truskett[2,3]

[1)] *Theoretical Division, Los Alamos National Laboratory*

[2)] *McKetta Department of Chemical Engineering, University of Texas at Austin, Austin, Texas 78712, USA*

[3)] *Department of Physics, University of Texas at Austin, 2515 Speedway, Austin, Texas 78712, United States*

While many physical processes are non-equilibrium in nature, the theory and modeling of such phenomena lag behind theoretical treatments of equilibrium systems. The diversity of powerful theoretical tools available to describe equilibrium systems has inspired strategies that map non-equilibrium systems onto equivalent equilibrium analogs so that interrogation with standard statistical mechanical approaches is possible. In this work, we revisit the mapping from the non-equilibrium random sequential addition process onto an equilibrium multi-component mixture via the replica method, allowing for theoretical predictions of non-equilibrium structural quantities. We validate the above approach by comparing the theoretical predictions to numerical simulations of random sequential addition.

[a)] Electronic mail: rjadrich@lanl.gov

## I. INTRODUCTION

Theoretical treatment of non-equilibrium problems represents an important and formidable challenge in the modeling of physical phenomena. Driven systems[1–7] and active matter[8–10] are examples of non-equilibrium processes of interest in the field of material science. Such systems also display a diverse array of complex phase transitions.[11–13]. Furthermore, the biological processes relevant to life are inherently non-equilibrium.[14,15] Non-equilibrium processes are more complex than their equilibrium analogs in that one must consider an ensemble of dynamical trajectories (instead of an ensemble of states) and the history of the trajectory is relevant.[14,15]

Despite the significance and sheer abundance of non-equilibrium systems, there is no comprehensive theoretical framework for their modeling. By contrast, for systems in equilibrium, a broad array of statistical mechanical tools has been developed. Examples of such tools include theories of the mean-field, re-normalization group, and liquid-state closure varieties.[16–22] In addition to the aforementioned theoretical tools, the relationships between thermodynamic variables and how such quantities relate to phase transitions are also well established.[16–23]

In this work, we wish to leverage the vast body of work on equilibrium statistical mechanics to better understand and describe non-equilibrium systems. One path forward in this regard is to formulate a thermodynamic framework for non-equilibrium problems; several works have formulated definitions of entropy for non-equilibrium systems, for instance.[24–26] Maximum caliber is a generalization of this idea, where the distribution of dynamic trajectories (instead of the distribution of states in equilibrium) is inferred from the maximum entropy principle.[27–29]

One intriguing alternative possibility for theoretically describing non-equilibrium phenomena is to discover an approximate mapping from the non-equilibrium process to an equivalent equilibrium system. For a subset of non-equilibrium problems characterized by the presence of quenched disorder (i.e., degrees of freedom not in thermal equilibrium but rather frozen in place), the replica method (also known as the replica trick) provides a path forward.[30–32] While the replica method gained recognition with its first applications to spin glasses,[33,34] a more complete appreciation of its power followed from seminal work of Giorgio Parisi wherein nonphysical complications stemming from the replica trick were resolved through a phenomena called replica symmetry breaking.[35–37] The establishment of replica symmetry breaking and the unique hierarchical structure for breaking the symmetry was a central aspect of the 2021 Nobel Prize in Physics. Despite the counterintuitive mathematics of the replica trick, which include creating $m$ copies (replicas) of the thermal degrees of freedom and then sending $m \to 0$, it has enabled the solution of complex spin glass problems, in some cases yielding provably exact results.[30–32,35–39]

In the past decade, the utility of the replica method has also been demonstrated in traditional structural glasses.[40–48] Unlike spin glasses, structural glasses do not have any imbued quenched disorder. Nevertheless, for hard spheres, the replica method enables the identification of glassy basins from the equilibrium fluid equation of state and the tracking of the glassy state as it approaches jamming upon compression.[40–43] This is a remarkable demonstration of the replica method's ability to handle what is, nominally, considered a non-equilibrium phenomenon using purely equilibrium statistical mechanics. The theory also yields the complexity, the analog of configurational entropy, which is a count of the number of glassy states, as well as a provocative prediction of an ideal glass, which is the densest amorphous glass packing and is akin to a disordered crystal. While the existence of the ideal glass is still debated, the replica method provides an extremely comprehensive and microscopic predictive theory of structural glasses and jamming.

In this work, we leverage the replica method to approximately map the non-equilibrium random sequential addition (RSA) process[49–57] onto an equilibrium problem. RSA is conceptually simple: one particle is added to a box in a random position and frozen in place. A second particle is added at a random position, subject to the constraint that it does not overlap with the first particle. This procedure of adding particles randomly, so long as they do not generate particle overlaps, is repeated iteratively until no more particles can be placed in the box. Since the entire history of the process influences the end result, the procedure must, practically, be repeated until statistics converge.[58] RSA has some interesting properties, including a terminal (also called saturation) density[49–52,54] beyond which the process cannot be continued and an unusual logarithmic form[51,54] of the contact peak of the radial distribution function (RDF). Furthermore, unlike equilibrium hard spheres, correlations between spheres differ depending on the time point at which they were added. RSA is also a canonical example of a sequential exclusion physical process. Processes in this general family have been used to model real-world phenomena such as traffic flow and cell

transport.[59,60]

Previous work has recognized the utility of the replica method for RSA. In one case, a free energy for RSA was derived in the grand canonical ensemble and fit to a rational function approximation in two-dimensions in an attempt to extract the terminal density.[61] Other work has focused on the extension of replica integral equation theory[62–64] to RSA.[61,65,66] We significantly extend this body of work in several important ways. First, we apply the replica method to develop an expansion for structural correlations instead of the free energy. This formulation allows for examination of hard sphere contact correlations as a function of the order in which they were added to the system, allowing for predictions on a per particle level. We also show results from one- to six-dimensions, and we provide a clear description of (and justification for) which graphical terms are included in the theory, providing a road map for further theoretical developments. Finally, the results of this work are timely with respect to recent replica theory developments in the structural glass community;[40–48] specifically, this body of work may shed light on the apparent lack of a terminal density in replica theories (including this one) for RSA.[65,66]

The remainder of the manuscript is organized as follows. In Section II, we describe the mapping between the RSA process and an equivalent equilibrium system via the replica method. We defer the bulk of the mathematical details to the Appendix. In Sect. III, we provide computational details for the RSA simulations and compare the results of the theory to RSA numerical simulations, where we show that the agreement between theory and simulation is very good. Finally, we conclude and provide an outlook in Sect. IV.

## II. THEORY

In order to map the RSA process onto an equilibrium system, we employ the replica method—a powerful mathematical tool that allows for the thermodynamic evaluation of systems possessing quenched (frozen) and thermal (ergodic) degrees of freedom.[30–32] Originally developed for spin glasses,[33,34] it provides a recipe for extracting properties of the real quenched disorder system from a fictive isomorphic system whereby the quenched degrees of freedom are treated on the same footing as the thermal analogs. As a relevant example, consider the multi-step process of equilibrating hard spheres at some finite number density ($\rho_1$), freezing these spheres in place, and then adding and equilibrating a second "batch" of hard spheres with density $\rho_2$ in the presence of the frozen spheres. The thermodynamic properties of this hybrid frozen/ergodic system are complicated and are not equivalent to a system of equilibrium hard spheres at density $\rho_1 + \rho_2$. This is where the replica method enters. The isomorphic equilibrium system can be thought of as a single copy of the frozen spheres in the presence of $m$ copies (replicas) of the mobile spheres.[62–64] Within a single copy, the particles of the mobile system mutually interact; however, particles in different replicas are non-interacting. The entire system (even the originally frozen spheres) is fully thermalized. The replica method is then used to compute the relevant thermodynamic quantities at integer values of $m$ and then $m$ is analytically continued to zero to recover the original quenched disorder system.

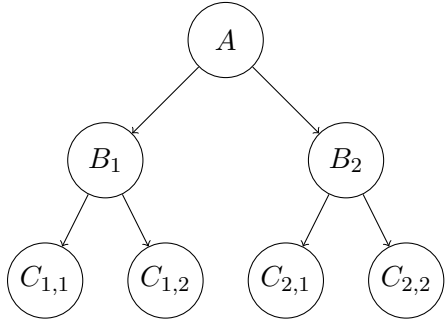

FIG. 1. Schematic for the interactions in a replicated system for $m_B$ and $m_C = 2$.

The RSA process is related to (but more complicated than) the simple example above, possessing an infinite hierarchical form of quenched disorder. Each addition is a quenched disorder problem, where the particles already placed in the box are frozen and the particle that is being added is thermalized. Therefore, the equivalent equilibrium system in the thermodynamic limit is an infinite-component mixture with a tiered structure that can be imagined as follows. The first particle added is represented as a single $A$ particle that interacts with $m_B$ copies of a single $B$ particle. The copies of $B$ are mutually pairwise non-interacting, but they all interact with the single $A$ particle. Similarly, each $B$ particle gets its own $m_C$ copies of a single $C$ particle. None of the $m_C \times m_B$ copies of the $C$ particles interact directly with each other, and they only directly interact with "their" $B$ copy. All $C$ replicas interact with the $A$ particle. This structure is repeated infinitely.[67] This interaction hierarchy is depicted graphically in Fig. 1 for $m_B, m_C = 2$, where the only particles that directly interact are connected by a contiguous pathway of downward-facing arrows.

Because the above system is fully thermalized and amounts to a multi-level Widom-

Rowlinson mixture,[68–70] we use liquid-state theory to compute the quantities of interest (here, the contact value of the radial distribution function $g(\sigma)$, where $\sigma$ is the hard core diameter) and the replica method is invoked to compute the values for the RSA process. As derived in Appendix A, the relationship between the RDF of the real system and the replicated system between spheres added at addition $\kappa$ out of a total of $n$ additions is

$$g_{\kappa,n}(r) = \lim_{m_n \to 0} \frac{\partial}{\partial m_n} \cdots \lim_{m_2 \to 0} \frac{\partial}{\partial m_2} m_2 m_3 \cdots m_n g_{\kappa,n}(r|\boldsymbol{m}) \qquad (1)$$

where $g_{\kappa,n}(r|\boldsymbol{m})$ is the partial RDF between particles at level $\kappa$ and $n$ in the replicated tree structure that are connected by a continuously descending path, $\boldsymbol{m} \equiv \{m_2, m_3, ..., m_n\}$ is the set of the number of replicated copies at each level, and $g_{\kappa,n}(r)$ is the real partial RDF between particles. This further simplifies to

$$g_{\kappa,n}(r) = g_{\kappa,n}(r|\boldsymbol{m} \to 0). \qquad (2)$$

Correlations between species not connected by a continuously descending path (sometimes called blocking correlations) also have a physical connection to the real RSA process, though it is more obscure. Such replica "blocking" correlations can provide the real correlations to the following example.[62–64] In the real RSA process, we can look across separate realizations of particle additions where we add particles identically up to some density and after which we follow different addition sequences. Particles added after the randomization step will be correlated across realizations, but only by virtue of their shared history. We do not pursue blocking correlations in this study and instead reserve their treatment for future work.

In formulating our theoretical approach, we pursue a virial expansion[19] of the replicated mixture for $g_{\kappa,n}(r|\boldsymbol{m})$ at contact according to standard liquid-state theory in terms of 2-, 3-, 4- body interactions.[71,72] For convenience, the standard liquid-state theory virial expansion is discussed in Appendix B within the context of this work. Non-zero contributions to the virial coefficients can be enumerated using graphs, as discussed in Appendix C. It can be shown that only pairs (triplets) of particles in the above equilibrium system which directly interact with each other contribute to the second (third) virial coefficient. The fourth-order virial coefficient is more complicated to compute because some of the interactions in the quartet of particles can be broken and still generate a finite contribution.[19,73–76] Beyond the fourth coefficient, the complexity grows rapidly; therefore, we truncate the expansion at fourth order. The final fourth order expansion, derived in Appendix C, after taking the

$m \to 0$ limit is

$$g_{\kappa,n}(\eta_\kappa, \eta_n) \approx 1 + Q_3 \eta_n + \left(Q_4 - \widetilde{Q}_4\right)\eta_n^2 + \widetilde{Q}_4 \eta_\kappa^2 \tag{3}$$

where $\eta_i \equiv v_D \sigma^D \rho_i$ and $\rho_i$ are the total volume fraction and number density, respectively, after the $i$th RSA addition, $v_D$ and $s_D$ are the volume and surface area of a unit sphere in $D$ dimensions, respectively, $\sigma$ is the hard sphere diameter, and the coefficients are

$$Q_3 \equiv \frac{3/2}{v_D(s_D/2)\sigma^{2D-1}} \left.\frac{\partial B_{a,b,c}}{\partial \sigma_{a,b}}\right|_{\sigma_{i,j}=\sigma} \tag{4}$$

$$Q_4 \equiv \frac{2}{v_D^2(s_D/2)\sigma^{3D-1}} \left.\frac{\partial B_{a,b,c,d}}{\partial \sigma_{a,b}}\right|_{\sigma_{i,j}=\sigma} \tag{5}$$

$$\widetilde{Q}_4 \equiv \frac{2}{v_D^2(s_D/2)\sigma^{3D-1}} \left.\frac{\partial B_{a,b,c,d}}{\partial \sigma_{a,b}}\right|_{\substack{\sigma_{i,j}=\sigma,\\ \sigma_{b,d}=0}} \tag{6}$$

where $B_{1,2,\ldots,n}$ are the standard species dependent virial coefficient from liquid state theory[19,73] and $\sigma_{a,b}$ are the diameters between species $a$ and $b$ in the virial coefficients. Numerical evaluation of the diameter derivatives with respect to the third and fourth virial coefficients are discussed in Appendices D-E.

| $D$ | $Q_3$ | $Q_4$ | $\widetilde{Q}_4$ | $Z_1$ | $Z_2$ | $A$ |
|---|---|---|---|---|---|---|
| 1 | 1 | 1 | -1/4 | 1/4 | 1/4 | 0 |
| 2 | 1.5640 | 2.1289 | -0.87808 | 0.56143 | 0.56201 | 0.43599 |
| 3 | 2.5000 | 4.5912 | -2.8857 | 1.1543 | 1.1908 | 1/2 |
| 4 | 4.0507 | 9.7181 | -9.2206 | 2.2763 | 2.1566 | -0.050720 |
| 5 | 6.6250 | 19.449 | -29.033 | 4.3824 | 3.8275 | -1.6250 |
| 6 | 10.910 | 34.164 | -90.631 | 8.3071 | 6.8133 | -4.9101 |

TABLE I. Values for the parameters of the contact RDF virial expansion ($Q_3$, $Q_4$, $\widetilde{Q}_4$) and the Carnahan-Starling (CS) corrected form ($Z_1$, $Z_2$, $A$) from one to six dimensions.

We further leverage liquid-state theory to attempt to correct the truncated expansion above. For hard spheres, the analogous virial expansion underpredicts the entropy loss as a function of density (i.e., the available space is over-predicted more dramatically with increasing density).[19] Part of the issue is that a virial series expansion is not rapidly convergent.[19,22,77–79] The Carnahan-Starling equation of state (CS-EOS) circumvents this difficulty by approximately re-summing the terms in the virial expansion as a geometric series that can be analytically evaluated, resulting in a nearly exact expression for the contact value (and all other thermodynamic properties) of the equilibrium hard-sphere liquid phase

$$g_{\text{eq}}(\eta) = \frac{1 - A\eta}{(1-\eta)^D} \qquad (7)$$

where $\eta = v_D \sigma^D \rho$ in the volume fraction, and $A$ has a simple analytical form for all dimensions.[22,43,77–79] The CS form is virtually exact over the entire fluid regime at all tested dimensions (and polydisperse mixtures) for equilibrium hard spheres.[43,80,81] We leverage the CS relation as an approximate way to also "re-sum" higher order effects for RSA via the following ansatz

$$g_{\kappa,n}(\eta_\kappa, \eta_n) \approx g_{\text{eq}}(\eta_n - Z_1 \eta_\kappa^2 + Z_2 \eta_n^2) \qquad (8)$$

where $Z_1$ and $Z_2$ are yet to be determined coefficients. We choose to set the unknown coefficients by forcing the series expansion of Eqn. 8

$$\begin{aligned} g_{\text{eq}}(\eta_n - Z_1 \eta_\kappa^2 + Z_2 \eta_n^2) =& 1 + (D - A)\eta_n + \\ & \left[\frac{1}{2}D(D - 2A + 1) + Z_2(D - A)\right]\eta_n^2 - \\ & Z_1(D - A)\eta_\kappa^2 + \cdots \end{aligned} \qquad (9)$$

to agree with that in Eqn. 3 for each term in density. By design, the lowest order density term from the CS relation recovers the third virial coefficient, hence our neglect of a scalar to multiply the $\eta_n$ term in Eqn. 8. Solving for equality of the quadratic density terms yields

$$Z_1 = \frac{\widetilde{Q}_4}{A - D} \qquad (10)$$

and

$$Z_2 = \frac{Q_4 - \widetilde{Q}_4}{D - A} - \frac{D(D - 2A + 1)/2}{D - A} \qquad (11)$$

As discussed in Sect. III, this approximate re-summed form (Eqn. 8) has a larger domain of validity than the low density expansion (Eqn. 3) alone. It also has a terminal density, by construction; however, it is far larger than the observed values. For example, our theory in 2D has a terminal density at $\eta = 0.714$, which is much larger than the known value of $\eta = 0.547$,[54] and the predictions do not improve with dimension. For convenience, tabulated values for $Q_3$, $Q_4$, $\widetilde{Q}_4$, $Z_1$, $Z_2$, and $A$ from one to six dimensions are provided in Table I.

From the partial radial distribution functions at contact for RSA, it is easy to compute the total radial distribution function at contact. As elaborated upon in Appendix F, the calculation is a straightforward double integral over the continuous sequential additions

$$g_{\rm rsa}(\eta) = \frac{2}{\eta^2} \int_0^\eta d\eta_\kappa \int_{\eta_\kappa}^\eta d\eta_n g_{\kappa,n}(\eta_\kappa, \eta_n) \tag{12}$$

We will use both the simple expansion (Eqn. 3) and the CS improved form (Eqn. 8) to compare to exact simulation results.

### III. RESULTS AND DISCUSSION

In this section, we validate predictions of the replica theory of this study by comparison to direct numerical simulations of the RSA process. In particular, we compare the theory and simulation contact values for both the total RDF and the partial RDFs. The partial RDFs are grouped on the basis of the order in which they are added to the simulation box, which is equivalent to the alphabetic labels for the equivalent equilibrium system described in Sect. II. To evaluate the accuracy of the above theory, direct simulations of the RSA process for systems in six different spatial dimensions were performed. Computational cost grows rapidly with increasing dimensionality, necessitating the use of cell lists to speed up the simulations. At each density, statistics for the contact value of the total RDF were found to be well converged after roughly $\mathcal{O}(10)$ separate realizations of a 10,000 particle simulation. A larger number $\mathcal{O}(100)$ of separate realizations were used to gather partial radial distribution functions in 3D. Our simulations allowed up to 1,000,000 insertion attempts before we stopped simulating. As dimensionality increases, approaching the terminal density becomes more difficult; as such, we do not get as close to the terminal density in higher dimensions (though this does not inhibit the validation of the theoretical framework). Virtually exact terminal densities, up to 8D, are known from a study using a more sophisticated algorithm aimed at probing the terminal density directly.[54]

Fig. 2 compares contact values of the total RDF computed in various ways for one- to six-dimensional hard-sphere RSA processes. The x-axis in each subplot of Fig. 2 extends only up to the terminal volume fraction for the RSA process as determined by prior simulations for that dimensionality.[54] The RSA simulation results are plotted as black open circles. For comparison, the dot-dashed teal lines show the CS-EOS contact values for equilibrium hard spheres. At lower to intermediate packing fractions, the contact value for RSA simulations is lower than for equilibrium hard spheres due to the lack of two-body correlations in the random insertion process. However, as the density increases, the RSA process runs out of free volume more quickly because there is no correlated motion or rearrangement, which results in the contact value swiftly increasing, eventually crossing over the equilibrium hard-sphere values.

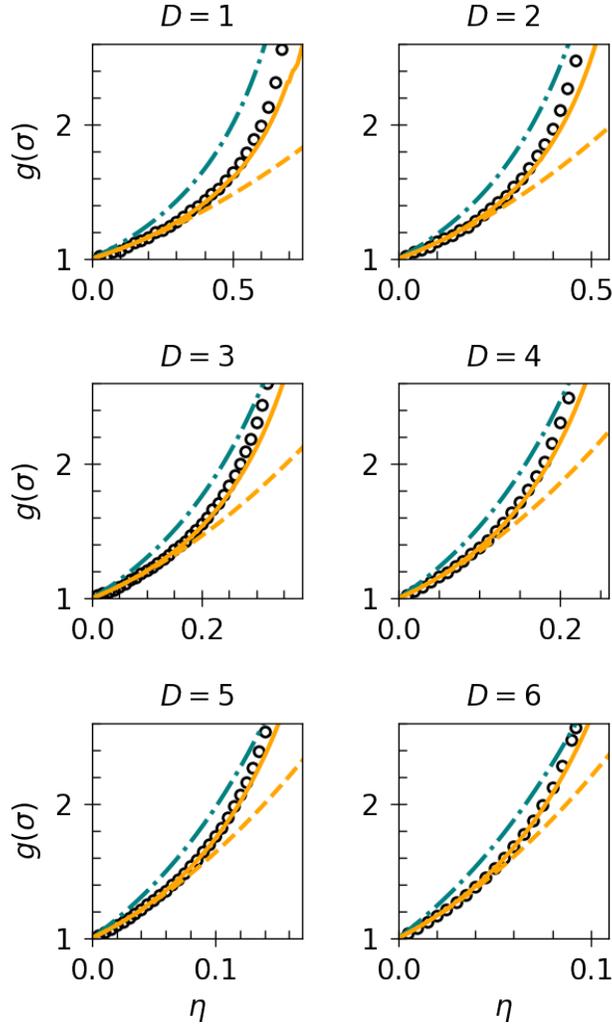

FIG. 2. Comparison of total radial distribution function at contact ($g(\sigma)$) for direct simulation of RSA (black open circles), the corrected (solid orange line) and uncorrected (dashed orange line) replica theory for RSA, and equilibrium hard spheres (dot dashed teal line) according to the CS contact value. Higher volume fractions (beyond the rightmost value of each plot) are inaccessible to RSA due to the presence of a terminal density and a jamming-like transition.

The theory derived in Sect. II follows in spirit from the liquid-state theory treatment of equilibrium hard spheres, where the approximations induced by truncating the virial expansion break down at higher packing fractions when higher-order correlations become more influential. This limitation also manifests in the uncorrected replica theory, which is plotted as short orange dashed lines in Fig. 2. As expected by analogy to equilibrium hard spheres, the agreement between theory and RSA simulation is good at low densities, but then breaks down as the packing fraction increases, with the theory underpredicting the contact value. Predictions do seem to improve with increasing dimensionality as one would suspect from the increasing ideality of equilibrium hard spheres with increasing dimension (i.e., at infinite $D$ only the second virial correction is required for equilibrium hard spheres).[82,83] The corrected (CS based) replica theory that approximately includes some of the missing higher-order terms, plotted as solid orange lines, is in better agreement with the simulation results. There is still some minor discrepancy at very high densities, possibly due to the missing RSA corrections at fifth- and higher order; discussion of other possible interpretations and future avenues for research along these lines is deferred to Sec. IV. Interestingly, it seems possible from Fig. 2 that the CS corrected (and uncorrected) replica theory may improve with increasing dimension, though further work is required to fully assess this. The CS corrected theory is probably more rapidly convergent than the uncorrected virial expansion though.

Because the uncorrected theory significantly differs from the simulated results at higher packing fractions and the corrective methodology that brings the results into alignment is somewhat *ad hoc*, we provide additional support that the theory is meaningfully capturing the physics of the RSA process by temporally decomposing the particles on the basis of the order in which they are added to the system. (In the equilibrium theory, the addition order corresponds to the "level" label described in Sect. II.) At $\eta = 0.15$ and $D = 3$, where the uncorrected theory, CS-corrected theory and simulations are all in excellent agreement, we compare $g(\sigma)$ of the temporal self- and cross-terms in Fig. 3a-b for the simulation and CS corrected theory, respectively. The particles are grouped into deciles: the first 10% of the particles added to the system, the second 10% of the particles added, and so on. In Fig. 3c, we plot $g(\sigma)$ from simulation between the first decile and the $n^{th}$ decile (going across the first row of the heat maps), and in Fig. 3d, we plot the $g(r)$ between the 10th decile and the $\kappa^{th}$ decile (going down the last column of the heat maps). Along with the RDFs, we show the contact value predicted by theory for each RDF as a horizontal dotted line. Across all panels of Fig. 3, we see near quantitative agreement between the simulated and theoretical contact values. The excellent agreement between theory and simulation in Fig. 3 provides strong evidence that the theoretical agreement with simulation is not fortuitous as it also captures the relatively fine-grain metric of temporally specific partial RDFs.

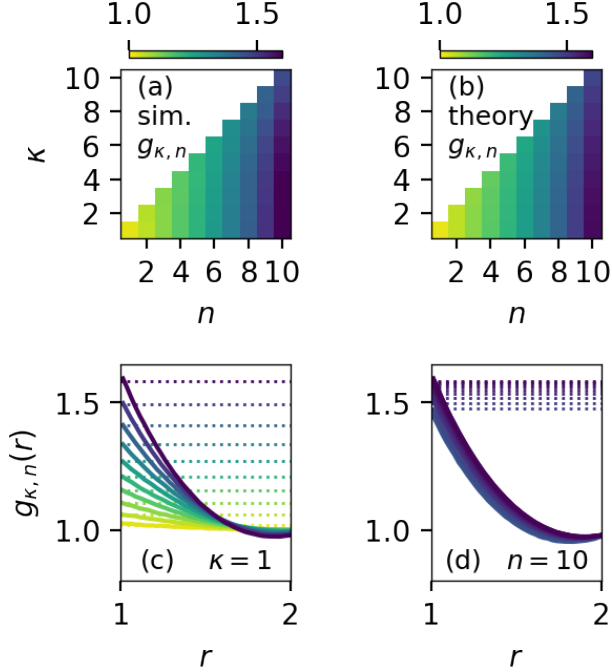

FIG. 3. Magnitude of the (a) simulated and (b) theoretically predicted partial RDF contact values (grouped into deciles based on the order of addition) for RSA in 3D at $\eta = 0.15$. (c) Solid lines are RDFs sweeping across $n$ at fixed $\kappa = 1$. Dotted lines of the same color indicate the theoretical prediction of the contact value. (d) The same as panel c, but sweeping across $\kappa$ for fixed $n = 10$.

Our convention in Fig. 3a-b is that $\kappa \leq n$, though the plot is symmetric diagonal. When $\kappa < n$, the $\kappa$ particles were frozen when the $n$ particles were added. For both the theory and the simulations, as $n$ increases for any value of $\kappa$, the contact value also increases noticeably; that is, particles that are added later in the RSA process have stronger correlations (in a two-body sense) with frozen particles. This trend is easily understood. As the simulation box fills up, it is increasingly likely that subsequent particles will be placed in close proximity to a frozen particle. As $\kappa$ increases (particularly for larger values of $n$), the contact value decreases, though the magnitude of the effect is much weaker. The origin of this effect is less obvious but can be imagined as follows. As the background density increases, there are a decreasing number of void spaces large enough to accommodate two particles. Therefore, at larger $\kappa$ values, particles that are added in close succession are actually less likely to be in close proximity to each other. Note that while we can rationalize the trends in Fig. 3 by leveraging physical intuition about the non-equilibrium RSA process, the same quantitative trends are present in the theoretical predictions as well.

## IV. CONCLUSIONS AND OUTLOOK

In this work, we developed a theory to describe the non-equilibrium RSA process by mapping RSA onto an isomorphic equilibrium system via the replica method. We validate the theory by comparing to direct simulations of RSA, showing good agreement between the RDFs at contact.

This work suggests several directions for future inquiry. The first is to reduce the degree of physics lost in the current theory by the truncation in the virial expansion. For instance, some other derivation, such as a Ree-Hoover expansion, could potentially have terms with complexity that grows less rapidly with the order of the expansion.[75,76] The second is to probe the infinite-dimension limit, where it may be possible to derive an exact expression via a full re-summation of ring diagrams[82,83] yielding a new high $D$ packing law based on RSA processes. Comparison of this scaling to the known result for the ideal glass and related jamming transition would be very interesting. Other potential extensions would be to modify the theory to account for additional complexities such as a time-dependent rate of addition in the random sequential process or particle size polydispersity. Ultimately, by building up a comprehensive theory for RSA, it might be possible to develop a comprehensive theoretical framework for all types of sequential exclusion processes.

There are also various questions that this theoretical framework invokes by way of analogy to the large body of work on the replica method as applied to structural glasses and jamming. First, the development of an expansion about the contact value would be informative to see if the unusual logarithmic form of the contact peak is recovered.[51,54] Replica theory for structural glasses has shown remarkable success in predicting the near contact behavior—obtaining nearly quantitative predictions in supercooled soft-sphere systems.[44,45] Second, it is known that one-step replica-symmetry breaking corresponds to the onset of configuration space fragmenting into separate basins (glassy states) in equilibrium fluids.[40–48] By analogy, it seems reasonable that replica symmetry breaking may be required to capture the more rapid growth of the contact value in RSA as the critical density is approached. The fragmentation of configuration space in RSA (if found) would likely be due to the previously quenched particles creating localized islands of configuration space for any new thermalized addition. Such a finding would support a more fundamental link between RSA insertion saturation and regular fluid jamming. Interestingly, the possible need for replica symmetry breaking is supported by liquid state replica symmetric integral equation theory studies of RSA wherein theory was found to vastly underestimate the RDF at densities near the saturation point and seemingly avoid any singularity entirely.[65,66] The same avoidance is found in a non-replica derived integral equation approach.[53] We note that our CS corrected theory has a singularity (by way of the denominator in Eqn. 7), but the resultant terminal density is far too high compared to the true values. Furthermore, the singularity in the theory is by construction and not emergent.

The general strategy employed in this paper is potentially applicable to certain other non-equilibrium processes as well, though they are more complex and emergent in nature. Diffusion limited aggregation[84] or colloidal gelation[85] may be approximated by repeated thermalization and quenching processes. Random Organization (RO), a non-equilibrium model for colloidal shearing, is another such process that also has an element of quenched disorder.[86–89] In RO, particles are randomly placed in a box. Particles that do not overlap are fixed, and particles with overlaps are active. The active particles move randomly: they could become fixed if they move such that they do not have overlaps, or they could remain active if the move does not relieve their original overlap or if their move generates a new particle overlap. (Similarly, inactive particles can become active if an active particle moves such that it overlaps with it.) The model has interesting phase behavior as a function of particle density and move size where, depending on these variables, simulations either relax to a quiescent state where all particles are inactive or to a steady state where active particles persist indefinitely. It would be interesting to see if the replica method could be adapted to the RO model, where particles can alternate between frozen and mobile (i.e., emergent quenched disorder). This current work in combination with future efforts could further the development of a rigorous theoretical treatment of non-equilibrium statistical mechanics.

**ACKNOWLEDGMENTS**

We acknowledge support from the Welch Foundation (Grant No. F-1696) and the Texas Advanced Computing Center (TACC) at The University of Texas at Austin. B.A.L. acknowledges support from Los Alamos National Laboratory through the Darleane Christian Hoffman Distinguished Postdoctoral Fellowship.

**Appendix A: Sequential replica trick**

The RSA process for hard spheres can be interpreted as a sequential addition, equilibration, and positional freezing (quenching) protocol.[61] At each step (indexed by $\kappa$), new particles interact with one another, and with the previous particles, via hard-core interaction potentials. For convenience of notation in this section, all energies (potentials, free energies, etc.) will be implicitly per unit of thermal energy, $k_\mathrm{B}T$, where $k_\mathrm{B}$ is Boltzmann's constant and $T$ is the temperature. Also for convenience, we assign book-keeping indices to the hard-core potential between particles added at steps $\kappa$ and $\gamma$ as $u_{\kappa,\gamma}(r|\sigma)$ where $r$ is the center-to-center distance between the two particles and $\sigma$ is the hard-core diameter. Thus, at step $\kappa$ the energy for the added particles is broken into a self term and a contribution from the new particles interacting with all the previously added, and now frozen, particles

$$U_\kappa(\boldsymbol{R}_\kappa|\boldsymbol{R}_{1:\kappa-1}) \equiv \sum_{i=1}^{N_\kappa}\sum_{j=i+1}^{N_\kappa} u_{\kappa,\kappa}(|\boldsymbol{r}_{i,\kappa}-\boldsymbol{r}_{j,\kappa}|) + \sum_{\gamma=1}^{\kappa-1}\sum_{i=1}^{N_\kappa}\sum_{j=1}^{N_\gamma} u_{\kappa,\gamma}(|\boldsymbol{r}_{i,\kappa}-\boldsymbol{r}_{j,\gamma}|) \quad (A1)$$

where $\boldsymbol{r}_{i,\kappa}$ is the position of the $i$th particle from the $\kappa$th addition and $\boldsymbol{R}_\lambda$ and $\boldsymbol{R}_{1:\lambda}$ are shorthand for the set of positions for addition $\lambda$ and 1 through $\lambda$ respectively. Thus, the equilibrium configurational probability distribution is

$$P_\kappa(\boldsymbol{R}_\kappa|\boldsymbol{R}_{1:\kappa-1}) = \exp[-U_\kappa(\boldsymbol{R}_\kappa|\boldsymbol{R}_{1:\kappa-1})]/Z_\kappa(\boldsymbol{R}_{1:\kappa-1}) \quad (A2)$$

where $Z_\kappa(\boldsymbol{R}_{1:\kappa-1}) \equiv \sum_{\boldsymbol{R}_\kappa} \exp[-U_\kappa(\boldsymbol{R}_\kappa|\boldsymbol{R}_{1:\kappa-1})]$ is the equilibrium configurational partition function. Furthermore, we will denote an average over $P_\kappa$ as $\langle\cdots\rangle_\kappa$. Only the configurational

contributions to the free energy and partition function need to be considered in this section as we seek only structural correlations.

To model the thermodynamics of a macroscopic RSA system that is self-averaging (i.e., thermodynamics does not depend on the realization of quenched disorder), we require the quench-disorder averaged configurational Helmholtz free energy for the whole RSA process

$$F_n \equiv \langle\langle \cdots \langle\langle -\ln Z_n \rangle_{n-1} \rangle_{n-2} \cdots \rangle_2 \rangle_1. \tag{A3}$$

This is a formidable quantity to compute theoretically as it is not amenable to treatment via the standard tools of equilibrium statistical mechanics. To map this problem onto the domain of equilibrium statistical mechanics we leverage the replica trick. First we define a replicated partition function

$$Z_n(\bm{m}) \equiv \langle\langle \cdots \langle\langle Z_n^{m_n} \rangle_{n-1}^{m_{n-1}} Z_{n-1}^{m_{n-1}} \rangle_{n-2}^{m_{n-2}} Z_{n-2}^{m_{n-2}} \cdots \rangle_2^{m_2} Z_2^{m_2} \rangle_1 Z_1 \tag{A4}$$

where $\bm{m} = [m_2, ..., m_n]$ are variables that can assume any real value. Defining the corresponding replicated free energy as

$$F_n(\bm{m}) \equiv -\ln Z_n(\bm{m}), \tag{A5}$$

it can be shown that the real free energy can be obtained from the replicated free energy via

$$F_n = \lim_{m_n \to 0} \frac{\partial}{\partial m_n} \cdots \lim_{m_2 \to 0} \frac{\partial}{\partial m_2} F_n(\bm{m}). \tag{A6}$$

For general $\bm{m}$, this does not simplify the calculation. However, for the special case of all positive integer $\bm{m}$, $Z_{\bm{m}}$ is the partition function for a complex, equilibrium non-additive mixture of spheres. This is easy to see from the form of Eqn. A4: (1) every average is multiplied by the corresponding partition function of equal power, effectively converting the average to a simple summation (integration) over the particle coordinates, and (2) non-additivity comes from the various powers of $m$ that effectively create $m$ non-interacting clones of the newly added particles at each level in the addition sequence. However, all of the clones interact identically with previously added particles.

Ultimately, the complex mixture can be described by a branched tree encoding the hierarchical relationship among species. At level $\kappa$ in the tree there are $m_2 m_3 \cdots m_\kappa$ nodes that represent non-interacting copies (replicas) of the set of particles added at stage $\kappa$ in the RSA process. Any one replica at level $\kappa$ has a parent node (replica) at level $\kappa - 1$ that is common to its $m_\kappa - 1$ siblings. Parent replicas interact with all of their descendants via a hard core repulsion. More specifically, any two replicas interact via a hard core if they are related via a continuously descending (or ascending) path in the tree; otherwise, they are non-interacting.

As we seek to predict the contact value of the radial distribution function, we require a relationship between the RDFs of the final added particles (level $n$) in RSA and that of earlier analogs at some arbitrary level $\kappa \leq n$. We obtain the relationship relating the real RDF to the replicated RDF by taking the functional derivative of Eqn. A6 with respect to $u_{\kappa,n}(|\boldsymbol{r}_1 - \boldsymbol{r}_2|)$ which yields

$$g_{\kappa,n}(r) = \lim_{m_n \to 0} \frac{\partial}{\partial m_n} \cdots \lim_{m_2 \to 0} \frac{\partial}{\partial m_2} m_2 m_3 \cdots m_n g_{\kappa,n}(r|\boldsymbol{m}) \tag{A7}$$

The replicated RDF is the radial distribution function between any pair of replicas at level $\kappa$ and the final level $n$ (replica symmetry has been assumed) that are connected by a continuously descending path (there are $m_2 m_3 \cdots m_n$ of them). After application of the derivatives and limits one finds

$$g_{\kappa,n}(r) = g_{\kappa,n}(r|\boldsymbol{m} \to 0). \tag{A8}$$

The "trick" is to derive an expression for the mixture in the case of all integer $\boldsymbol{m}$ and assume that this can be continued to all real values of $\boldsymbol{m}$.

**Appendix B: General density expansion for the mixture contact value**

Working in the canonical (NVT) ensemble, it is straightforward to show that the contact value between species $a$ and $b$ ($g_{a,b}(\sigma_{a,b})$) is given by

$$g_{a,b}(\sigma_{a,b}) = \frac{1}{(2-\delta_{a,b})\rho x_a x_b (s_D/2)\sigma_{a,b}^{D-1}} \frac{\partial f}{\partial \sigma_{a,b}} \tag{B1}$$

where $f$ is the excess Helmholtz free energy per particle and thermal energy, $\delta_{a,b}$ is the Kronecker delta, $x_a$ is the particle (mole) fraction of species $a$, $s_D$ is the surface area of a unit D-sphere, $\sigma_{a,b} = \sigma_{b,a}$ is the cross-diameter between species $a$ and $b$, and $\rho$ is the total number density.[71,72] To obtain an expansion in density we leverage the standard virial expansion

$$f = \sum_{i=1}^{n_C} x_i \ln x_i + \ln\rho - 1 + \sum_{i=1}^{\infty} \frac{\rho^i}{i} B_{i+1} \tag{B2}$$

where $B_i$ is the $i^{\text{th}}$ virial coefficient.[19,71,73,74] Substituting Eqn. B2 into Eqn. B1 yields

$$g_{a,b}(\sigma_{a,b}) = \frac{1}{(2-\delta_{a,b})x_a x_b (s_D/2)\sigma_{a,b}^{D-1}} \sum_{i=1}^{\infty} \frac{\rho^{i-1}}{i} \frac{\partial B_{i+1}}{\partial \sigma_{a,b}}. \tag{B3}$$

The composition dependence of the virial coefficients is apparent from the following decomposition into the species dependent VCs

$$B_i \equiv \sum_{\alpha_1=1}^{n_\text{C}} \sum_{\alpha_2=1}^{n_\text{C}} \cdots \sum_{\alpha_i=1}^{n_\text{C}} x_{\alpha_1} x_{\alpha_2} \ldots x_{\alpha_i} B_{\alpha_1,\alpha_2,\ldots,\alpha_i} \tag{B4}$$

where $n_\text{C}$ is the number of components.[19,71,72] Eqn. B3 requires the derivative of Eqn. B4 with respect to $\sigma_{a,b}$. Taking the derivative and collecting identical terms via the permutation symmetry of the species labels yields

$$\frac{\partial B_i}{\partial \sigma_{a,b}} = \frac{i(i-1)}{2}(2-\delta_{a,b})x_a x_b \sum_{\alpha_3=1}^{n_\text{C}} \cdots \sum_{\alpha_i=1}^{n_\text{C}} x_{\alpha_3}\ldots x_{\alpha_i} \frac{\partial B_{a,b,\alpha_3,\ldots,\alpha_i}}{\partial \sigma_{a,b}}. \tag{B5}$$

Substituting Eqn. B5 into Eqn. B3 yields

$$g_{a,b}(\sigma_{a,b}) = \frac{1}{(s_D/2)\sigma_{a,b}^{D-1}} \sum_{i=1}^{\infty} \frac{(i+1)}{2} \sum_{\alpha_3=1}^{n_\text{C}} \cdots \sum_{\alpha_i=1}^{n_\text{C}} \rho_{\alpha_3}\ldots\rho_{\alpha_{i+1}} \frac{\partial B_{a,b,\alpha_3,\ldots,\alpha_{i+1}}}{\partial \sigma_{a,b}}. \tag{B6}$$

The virial coefficient derivatives are related to the standard Mayer-f function $f(r)$ of equilibrium statistical mechanics and can be expressed in a convenient graphical form.[19,73,74] For hard spheres, $f(r)$ depends only on the core diameter ($\sigma$) and is trivially related to Heaviside step function, $H(r)$, via $f(r|\sigma) = -H(\sigma - r)$. For succinctness, we define the additional function $\tilde{f}(r|\sigma) \equiv \partial f(r|\sigma)/\partial \sigma$ which is related to the Dirac delta function, $\delta(r)$, via $\tilde{f}(r|\sigma) = -\delta(r - \sigma)$. Graphical expressions can be defined using these two functions. The second and third order terms are fairly simple,

$$\frac{\partial B_{a,b}}{\partial \sigma_{a,b}} = -\frac{1}{2} \; \text{\textcircled{a}}\cdots\text{\textcircled{b}} \tag{B7}$$

and

$$\frac{\partial B_{a,b,c}}{\partial \sigma_{a,b}} = -\frac{1}{3} \; \begin{array}{c}\text{\textcircled{c}}\\ \triangle \\ \text{\textcircled{a}}\cdots\text{\textcircled{b}}\end{array} \tag{B8}$$

where each graph represents an integrated product of $f(r|\sigma)$ functions (solid bond) and one $\tilde{f}(r|\sigma)$ (dashed bond) where the integration is over a Cartesian coordinate associated with each node specifying a specific pair of species.[19,73,74] Specifically, the third order graph in Eqn. B8 is formally $\propto \int \int \int d\boldsymbol{r}_a d\boldsymbol{r}_b d\boldsymbol{r}_c \tilde{f}(r_{a,b}|\sigma_{a,b}) f(r_{a,c}|\sigma_{a,c}) f(r_{b,c}|\sigma_{b,c})$. Importantly, if any

bond (f-function) vanishes (i.e., corresponds to non-interacting species pair) then the whole graph vanishes. This property will be key to identifying the finite species contributions from the replica tree and is particularly relevant starting at fourth order. Specifically, the fourth order term is more complex,[19,73,74] possessing multiple graphs with varying degrees of connectivity:

$$\frac{\partial B_{a,b,c,d}}{\partial \sigma_{a,b}} = -\frac{1}{8}\left[\, 2\,\Box + 5\,\boxtimes' + \boxtimes \,\right] \tag{B9}$$

where

$$2\,\Box = \Box_1 + \Box_2 \tag{B10}$$

and

$$5\,\boxtimes' = \boxtimes'_1 + \boxtimes'_2 + \boxtimes'_3 + \boxtimes'_4 + \boxtimes'_5 \tag{B11}$$

As some of the graphs possess broken bonds, they can support "cross replica" contributions to the density expansion, as they are non-interacting.

**Appendix C: Density expansion of the replica tree mixture contact value**

**1. Graphical description of replicated mixture interactions**

Applied to the replicated mixture, the summation in Eqn. B5 extends over all of the species described by the RSA replica tree. There is an infinite number of combinations to consider; however, by adopting replica symmetry (the assumption that any group of replicas with the same hierarchical relationship in the replica tree posses the same statistical correlations), the summation can be reduced to a sum over a finite number of realizable hierarchical relationships among $i$ species, each with a weighting that counts the number of equivalent possibilities. The various relationships can be summarized by an abbreviated graphical notation.

The second and third order virial coefficients (Eqns. B7 and B8, respectively) are composed of a single fully connected graph and can thus not support any non-interacting species pairs. As such, the only hierarchical relationship that is compatible is all species in a single descending path (and thus fully interacting). The continuously descending second and third relationships are expressed as

$$\mathcal{P}^{(2)} \equiv \begin{array}{c} \text{\textcircled{1}} \\ \downarrow \\ \text{\textcircled{$\kappa$}} \\ \downarrow \\ \text{\textcircled{$n$}} \end{array} \quad , \quad \mathcal{P}^{(3)}_\Gamma \equiv \begin{array}{c} \text{\textcircled{1}} \\ \downarrow \\ \boxed{\text{\textcircled{$\kappa$}} \downarrow \text{\textcircled{$\alpha_3$}}} \\ \downarrow \\ \text{\textcircled{$n$}} \end{array} \quad , \tag{C1}$$

where $\Gamma$ indicates the position of node $\kappa$ along the primary backbone relative to any "summed" nodes within the rectangular shaded "plate". The second virial term has no summed nodes so there is only one graph (hence the lack of the plate notation) whereas for the third virial graph $\kappa$ can come before or after the summed $\alpha_3$ node. Both graph sets in Eqn. C1 represent a primary backbone in the replica tree which is just one of the $m_2 m_3 \cdots m_n$ continuously descending paths selected by a specific choice of replicas at levels $\kappa$ and $n$. All replicas along the primary path fully interact with one another via just hard-sphere interactions. Things become a bit more complicated at the fourth virial level with the allowed graphs

$$\mathcal{P}^{(4)}_\Gamma \equiv \begin{array}{c} \text{\textcircled{1}} \\ \downarrow \\ \boxed{\text{\textcircled{$\kappa$}} \downarrow \text{\textcircled{$\alpha_3$}} \downarrow \text{\textcircled{$\alpha_4$}}} \\ \downarrow \\ \text{\textcircled{$n$}} \end{array} \quad , \quad \mathcal{B}^{(4)}_\Gamma \equiv \begin{array}{c} \text{\textcircled{1}} \\ \downarrow \\ \boxed{\text{\textcircled{$\kappa$}} \downarrow \text{\textcircled{$\alpha_3$}}} \\ \downarrow \searrow \\ \text{\textcircled{$n$}} \quad \alpha_4 \end{array} \tag{C2}$$

where the first is just the primary path graph relevant at all virial levels and the second is a new branched graph with a single dangling species that resides one step off of the main path. Replicas on a branch do not interact with the those on the primary path that come after the branch point. The first three graphs ($\mathcal{P}_\Gamma$) correspond to replicas that fully interact with one another (just hard spheres) and the latter two graphs ($\mathcal{B}_\Gamma$) have one pair of replicas that do not interact. Examples of graphs that do not contribute at the fourth virial level are

$$\begin{array}{c} \text{\textcircled{1}} \\ \downarrow \\ \boxed{\text{\textcircled{$\kappa$}} \searrow \text{\textcircled{$\alpha_3$}} \quad \alpha_4} \\ \downarrow \\ \text{\textcircled{$n$}} \end{array} \qquad \begin{array}{c} \text{\textcircled{1}} \\ \downarrow \\ \text{\textcircled{$\kappa$}} \\ \swarrow \downarrow \searrow \\ \alpha_3 \quad \text{\textcircled{$n$}} \quad \alpha_4 \end{array} \tag{C3}$$

as they have too many "broken" interactions between any one species. Specifically, $\alpha_4$ and $n$ have two species they do not interact with in the first and second graph types, respectively.

The fourth virial coefficient can support at most one disconnect for any of the species. Each higher order virial coefficient can support one more break, allowing for more complicated relationships. Finally, graphs with branches more than one node deep are irrelevant as they vanish in the $\bm{m} \to 0$ limit of RSA, as discussed below.

## 2. Density expansion

Using the results of the previous section, we can now calculate the contact value between particles added at different times during the RSA process. Replica symmetry is assumed at every step of replication in the tree, which is equivalent to assuming that any group of replicas with the same hierarchical relationship in the replica tree posses the same statistical correlations. To highlight the contributions from the specific contributions we "re-sum" terms in Eqn. B6 according to the various graphically described contributions in Eqns. C1 and C2 yielding

$$
\begin{aligned}
g_{\kappa,n}(\bm{m}) = {} & 1 + Q_3 \left( \sum_{\alpha_3=\kappa}^{n} \Delta\eta_{\alpha_3} + \sum_{\alpha_3=1}^{\kappa} \Delta\eta_{\alpha_3} \right) \\
& + 2Q_4 \left( \sum_{\alpha_3=\kappa}^{n} \sum_{\alpha_4=\alpha_3}^{n} \Delta\eta_{\alpha_3}\Delta\eta_{\alpha_4} + \sum_{\alpha_3=1}^{\kappa} \sum_{\alpha_4=\kappa}^{n} \Delta\eta_{\alpha_3}\Delta\eta_{\alpha_4} + \sum_{\alpha_3=1}^{\kappa} \sum_{\alpha_4=\alpha_3}^{\kappa} \Delta\eta_{\alpha_3}\Delta\eta_{\alpha_4} \right) \\
& + 2\widetilde{Q}_4 \left( \sum_{\alpha_3=\kappa}^{n} \sum_{\alpha_4=\alpha_3+1}^{n} (m_{\alpha_4}-1)\Delta\eta_{\alpha_3}\Delta\eta_{\alpha_4} + \sum_{\alpha_3=1}^{\kappa} \sum_{\alpha_4=\kappa+1}^{n} (m_{\alpha_4}-1)\Delta\eta_{\alpha_3}\Delta\eta_{\alpha_4} \right) + \cdots
\end{aligned} \quad \text{(C4)}
$$

where i) $Q_i$ are defined by Eqns. 4-6; ii) we have changed from number density ($\rho$) to volume fraction ($\eta$) and recognized the replica species densities in Eqn. B6 correspond to incremental densities added in the RSA process (hence the $\Delta$); iii) the factors of two account for permuting $\alpha_3$ and $\alpha_4$; iv) the factor of $m_{\alpha_4} - 1$ comes from the dangling $\alpha_4$ leaf in $\mathcal{B}_\Gamma^{(4)}$ of Eqn. C2 that is one removed from the primary descending path. Cases where $\alpha_4$ is two or more deep vanish in the limit $\bm{m} \to 0$ as multiplicative factors of $m$ get accrued that are not offset by any finite value. Using the definition of the total volume fraction $\eta_n \equiv \sum_{\alpha=1}^{n} \Delta\eta_\alpha$ further simplification yields

$$g_{\kappa,n} \equiv g_{\kappa,n}(\boldsymbol{m} \to 0) = 1 + Q_3(\eta_n - \eta_1)$$
$$+ 2Q_4\left(\eta_n(\eta_n - \eta_1) - \sum_{\alpha=1}^{n} \Delta\eta_\alpha \eta_\alpha\right) \quad (C5)$$
$$- 2\widetilde{Q}_4\left(\eta_n(\eta_n - \eta_\kappa) - \sum_{\alpha=\kappa}^{n} \Delta\eta_\alpha \eta_{\alpha+1} + (\eta_\kappa - \eta_1)(\eta_n - \eta_{k+1})\right) + \cdots$$

The result in Eqn. C5 does not assume anything about how many additions are performed or what is the size of each increment. In this work we limit our study to the constant rate addition of infinitesimal amounts characteristic of what is typically referred to as random sequential addition (though recognizing it is a subset of a family of processes). Setting $\Delta\eta_\alpha = \Delta\eta$, using $\eta_1 = \Delta\eta \to 0$, and summing the remaining sequence dependent terms in Eqn. C5 yields

$$g_{\kappa,n}(\eta_\kappa, \eta_n) \equiv g_{\kappa,n} = 1 + Q_3\eta_n + Q_4\eta_n^2 - \widetilde{Q}_4(\eta_n^2 - \eta_\kappa^2) + \cdots \quad (C6)$$

providing an exact low density expansion for the structural correlations between the sets of particles added in the RSA process. A slightly regrouped form of this equation is shown as Eqn. 3.

**Appendix D: Diameter derivatives of the fully-interacting composition dependent virial coefficients**

The main text requires derivatives of the form $\partial B_{a,b,\alpha_3,...,\alpha_i}/\partial \sigma_{a,b}$ evaluated at the point of all equivalent diameters. Using $\{...\}$ to denote a set, the point of equivalent diameters formally means $\sigma_{\kappa,\gamma} = \sigma$ for $\kappa, \gamma \in C(\{a, b, \alpha_3, ..., \alpha_i\})$ where $C(\{...\})$ generates the pair combinations of the entries in an arbitrary set, and $\sigma$ is the single desired diameter. For brevity, we will sometimes use the shorthand notation $\{\sigma_{\kappa,\gamma}\} = \sigma$ to indicate the aforementioned conditions.

The first step in the derivation is to take the total diameter derivative of the mixture virial coefficient with respect to $\sigma$

$$\frac{dB_{a,b,\alpha_3,...,\alpha_i}}{d\sigma} = \sum_{\kappa,\gamma \in C(\{a,b,\alpha_3,...,\alpha_i\})} \frac{\partial B_{a,b,\alpha_3,...,\alpha_i}}{\partial \sigma_{\kappa,\gamma}} \frac{\partial \sigma_{\kappa,\gamma}}{\partial \sigma} \quad (D1)$$

Under the condition that $\partial \sigma_{\kappa,\gamma}/\partial \sigma = 1$ and equivalent diameters, $\{\sigma_{\kappa,\gamma}\} = \sigma$, the quantity in Eq. D1 is trivially related to the total diameter derivative of the analogous hard sphere virial coefficient for monodisperse hard spheres of diameter $\sigma$ ($B_i^{\text{HS}}$) as

$$\left. \frac{dB_{a,b,\alpha_3,\ldots,\alpha_i}}{d\sigma} \right|_{\substack{\{\sigma_{\kappa,\gamma}\}=\sigma \\ \partial\sigma_{\kappa,\gamma}/\partial\sigma=1}} = \frac{dB_i^{\text{HS}}}{d\sigma} \qquad (D2)$$

This equivalence is a consequence of the definitions of partial and total derivatives and the functional form of $B_{a,b,\alpha_3,\ldots,\alpha_i}$ and $B_i^{\text{HS}}$—which are identical apart from explicit species labels and corresponding diameter labels. In the same limit, the right hand side of Eqn. D1 yields $i(i-1)/2$ identical terms, which combined with Eqn. D2 yields

$$\frac{dB_i^{\text{HS}}}{d\sigma} = \frac{i(i-1)}{2} \left. \frac{\partial B_{a,b,\alpha_3,\ldots,\alpha_i}}{\partial \sigma_{a,b}} \right|_{\{\sigma_{\kappa,\gamma}\}=\sigma} \qquad (D3)$$

Furthermore, as $B_i^{\text{HS}}$ is of the form $B_i^{\text{HS}} \propto \sigma^{D(i-1)}$ we have

$$\frac{dB_i^{\text{HS}}}{d\sigma} = \frac{D(i-1)}{\sigma} B_i^{\text{HS}} \qquad (D4)$$

Conveniently, analytic results exist for $B_i^{\text{HS}}$ for $i \leq 4$ in many dimensions, making it easy to evaluate $\partial B_{a,b,\alpha_3,\ldots,\alpha_i}/\partial \sigma_{a,b}$ for $\{\sigma_{\kappa,\gamma}\} = \sigma$ via Eqns. D3 and D4.[90]

**Appendix E: Diameter derivative of the singly-non-interacting composition dependent fourth virial coefficient**

As discussed in the main text, the first term that captures contributions involving unrelated species in the replica tree is that coming from the fourth virial coefficient. Specifically, we must evaluate $\partial B_{a,b,c,d}/\partial \sigma_{a,b}$ at the point $\sigma_{a,b} = \sigma_{a,c} = \sigma_{a,d} = \sigma_{b,c} = \sigma_{c,d} = \sigma$ and $\sigma_{b,d} = 0$. The core conditions make many of the Mayer-f graph contributions in Eqn. B9 vanish, requiring the evaluation of only

$$\left. \frac{\partial B_{a,b,c,d}}{\partial \sigma_{a,b}} \right|_{\sigma_{b,d}=0} = -\frac{1}{8} \left[ \begin{array}{c} \text{\scriptsize graph} \end{array} + \begin{array}{c} \text{\scriptsize graph} \end{array} \right] \qquad (E1)$$

where each solid line indicates a Mayer-f function between the species, a dashed line indicates a derivative of the Mayer-f function with respect to the particle diameter. Referring to the left and right graph as $\mathcal{G}_4$ and $\mathcal{G}_5$ respectively we can explicitly write them in a unified form as

$$\mathcal{G}_\lambda \equiv \int d\boldsymbol{r}_{d,a} \int \boldsymbol{r}_{b,a} \int \boldsymbol{r}_{c,a} f_{d,a}(r_{d,a}) F^{(\lambda)}_{c,a}(r_{c,a}) \frac{\partial f_{b,a}(r_{b,a})}{\partial \sigma_{b,a}} \\ \times f_{c,d}(|\boldsymbol{r}_{c,a} - \boldsymbol{r}_{d,a}|) f_{b,c}(|\boldsymbol{r}_{b,a} - \boldsymbol{r}_{c,a}|) \quad (E2)$$

where

$$F^{(\lambda)}_{c,a}(r_{c,a}) \equiv \delta_{\lambda,4} + \delta_{\lambda,5} f_{c,a}(r_{c,a}). \quad (E3)$$

$D = 1$ is a special case with the closed form results $\mathcal{G}_4 = 4\sigma^2$ and $\mathcal{G}_5 = -3\sigma^2$. For the case $D \geq 2$ we will show that Eqn. E2 can be reduced down from a $3 \times D$ dimensional integral to that of a double integral which is easily evaluated by quadrature.

To make progress we will utilize a few simplifications. First, the derivative of the Mayer-f function for hard spheres is simply a Dirac delta function

$$\frac{\partial f_{\kappa,\gamma}(r_{\kappa,\gamma})}{\partial \sigma_{\kappa,\gamma}} = -\delta(r_{\kappa,\gamma} - \sigma_{\kappa,\gamma}). \quad (E4)$$

Secondly we will convert D-dimensional Cartesian integrals to analogous D-dimensional spherical coordinate based integrals

$$\int d\boldsymbol{r} H(r) = s_D \int_0^\infty dr\, r^{D-1} H(r) \quad (E5)$$

and

$$\int d\boldsymbol{r} H(r,\theta) = s_{D-1} \int_0^\infty dr\, r^{D-1} \int_0^\pi d\theta \sin^{D-2}\theta H(r,\theta) \quad (E6)$$

where $\theta$ is an arbitrary angle, $s_D$ is the surface area of a unit $D$-dimensional sphere, and $H(r)$ and $H(r,\theta)$ are arbitrary radial and polar functions respectively.[74] Finally, we will utilize the definition of the vector norm, or equivalently the law of cosines, to write

$$|\boldsymbol{r}_1 - \boldsymbol{r}_2|^2 = r_1^2 + r_2^2 - r_1 r_2 \cos\theta \quad (E7)$$

where here $\theta$ is the angle between vectors $\boldsymbol{r}_1$ and $\boldsymbol{r}_2$. Using these simplifications we arrive at

$$\mathcal{G}_\lambda = \int d\boldsymbol{r}_{d,a} \int d\boldsymbol{r}_{c,a} f_{d,a}(r_{d,a}) F^{(\lambda)}_{c,a}(r_{c,a}) f_{c,d}(|\boldsymbol{r}_{c,a} - \boldsymbol{r}_{d,a}|) Z_{b,c}(r_{c,a}) \quad (E8)$$

where

$$Z_{b,c}(r_{c,a}) \equiv -s_{D-1} \sigma_{b,a}^{D-1} \int_0^\pi d\theta \sin^{D-2}\theta f_{b,c}\left(\sqrt{\sigma_{b,a}^2 + r_{c,a}^2 - \sigma_{b,a} r_{c,a} \cos\theta}\right). \quad (E9)$$

The integral in Eqn. E9 can be evaluated analytically for $\sigma_{i,j} = \sigma$, which after dropping species labels yields

$$Z(r) = -s_{D-1}\sigma^{D-1}\left[\frac{\sqrt{\pi}\Gamma\left(\frac{D}{2}-\frac{1}{2}\right)}{2\Gamma(D/2)} - {}_2F_1\left(\frac{1}{2},\frac{3-D}{2};\frac{3}{2};\frac{r^2}{4\sigma^2}\right)\frac{r}{2\sigma}\right] \tag{E10}$$

where ${}_2F_1(a,b;c;z)$ is the ordinary Hypergeometric function and $\Gamma(x)$ is the Gamma function. One final simplification can be achieved by leveraging the properties of convolutions and Fourier transforms to arrive at the two dimensional integral

$$\begin{aligned}\mathcal{G}_\lambda =& T_\lambda s_D s_{D-1}\pi^{1/2}\Gamma(D/2-1/2)\sigma^D \times \\ & \int_0^\infty dk\, k^{D-1}\frac{J^2_{D/2}(k\sigma)}{k^D}\int_0^{R_\lambda} dr\, r^{D-1}\frac{J_{D/2-1}(kr)}{(kr/2)^{D/2-1}}Z(r)\end{aligned} \tag{E11}$$

where

$$T_\lambda \equiv \begin{cases} -1 & \lambda = 4 \\ 1 & \lambda = 5 \end{cases} \tag{E12}$$

and

$$R_\lambda \equiv \begin{cases} \infty & \lambda = 4 \\ \sigma & \lambda = 5 \end{cases} \tag{E13}$$

and $J_v(x)$ is the Bessel function of the first kind of order $v$.

**Appendix F: Random sequential addition structural history integral**

The random sequential addition (RSA) process of the main text can be viewed as a sequence of $n$ steps whereby hard spheres are added to a volume $V$ until reaching the total density $\rho$ via increments of $\Delta\rho_i$, where $1 \leq i \leq n$, and are forevermore frozen in place for any subsequent additions. Defining $g_{i,j}(r)$ as the radial distribution function between particles added during addition $i$ and $j$ respectively, it is trivial to compute the total radial distribution function via the density weighted average

$$\rho^2 g(r) \equiv \sum_{i=1}^n \sum_{j=1}^n \Delta\rho_i \Delta\rho_j g_{i,j}(r) \tag{F1}$$

We will find it useful to rewrite Eqn. F1 such that self and cross-terms are separate

$$\rho^2 g(r) \equiv \sum_{i=1}^{n} \Delta\rho_i^2 g_{i,i}(r) + 2 \sum_{i=1}^{n} \sum_{j=i+1}^{n} \Delta\rho_i \Delta\rho_j g_{i,j}(r) \tag{F2}$$

In the continuous addition limit of RSA: $\Delta\rho_i = \Delta\rho = \rho/n$ and $n \to \infty$, the first term in Eqn. F2 vanishes yielding

$$\rho^2 g(r) = 2 \int_0^{\rho} d\rho_1 \int_{\rho_1}^{\rho} d\rho_2 \, g_{1,2}(r|\rho_1, \rho_2) \tag{F3}$$

where $g(r|\rho_1, \rho_2)$ is the radial distribution function between particles added when the density reaches $\rho_1$ and $\rho_2 > \rho_1$ during the RSA process.